\documentclass[12pt,preprint]{aastex}
\slugcomment{2013.3 v5}

\shorttitle{Wind braking of magnetars} \shortauthors{Tong, Xu, Song & Qiao}

\begin{document}

\title{Wind braking of magnetars}

\author{H. Tong\altaffilmark{1,3}, R. X. Xu\altaffilmark{2}, L. M. Song\altaffilmark{3} and G. J. Qiao\altaffilmark{2}
}

\altaffiltext{1}{Xinjiang Astronomical Observatory, Chinese Academy of Sciences, Urumqi, Xinjiang 
830011, China; tonghao@xao.ac.cn}
\altaffiltext{2}{KIAA and School of Physics, Peking University, Beijing 100871, China}
\altaffiltext{3}{Institute of High Energy Physics, Chinese Academy of Sciences, Beijing 100049, China}


\begin{abstract}
Considering recent observations challenging the traditional magnetar model, we explore the wind
braking of magnetars.
There is evidence for strong multipole magnetic fields in active magnetars, 
but the dipole field inferred from spin down measurements may be strongly biased by a particle wind.
Recent challenging observations of magnetars may be explained naturally in the wind braking scenario:
(1) The supernova energies of magnetars are of normal value;
(2) The non-detection in {\it Fermi} observations of magnetars; (3) The problem posed by the low-magnetic field
soft gamma-ray repeaters; (4) The relation between magnetars and high magnetic field pulsars;
(5) A decreasing period derivative during magnetar outbursts.
Transient magnetars with $L_{\rm x}<-\dot{E}_{\rm rot}$ may still be magnetic dipole braking.
This may explain why low luminosity magnetars are more likely to have radio emissions.
A strong reduction of dipole magnetic field is possible only when the particle wind 
is very collimated at the star surface. A small reduction of dipole magnetic field may result from
detailed considerations of magnetar wind luminosity. 
In the wind braking scenario, magnetars are neutron stars with strong multipole field. 
For some sources, a strong dipole field may be no longer needed.
A magnetism-powered pulsar wind nebula will be one of the consequences of
wind braking. For a magnetism-powered pulsar wind nebula, we should see a correlation between
the nebula luminosity and the magnetar luminosity. 
Under the wind braking scenario, a braking index smaller than three is expected.
Future braking index measurement of a magnetar may tell us whether magnetars are wind braking or
magnetic dipole braking.

\end{abstract}


\section{Introduction}

Anomalous X-ray pulsars (AXPs) and soft gamma-ray repeaters (SGRs) are magnetar
candidates, i.e., neutron stars powered by strong magnetic field decay (Thompson \& Duncan 1995, 1996).
In studying them,
the assumption of magnetic dipole braking is often employed (Duncan \& Thompson 1992; Kouveliotou et al. 1998).
However, the magnetic dipole braking mechanism is originally designed for rotation-powered pulsars.
Since both the persistent and burst emissions of magnetars are from a different energy reservoir
(magnetic energy instead of rotational energy), it is possible that they  have a different
braking mechanism, e.g., wind braking (Harding et al. 1999; Thompson et al. 2000).

A strong dipole magnetic field obtained by assuming magnetic dipole braking is often taken as
confirmation of a neutron star's magnetar nature ($B_{\rm dip}> B_{\rm QED}=4.4\times 10^{13} \, \rm G$, Kouveloitou et al. 1998). However,
the magnetic dipole braking assumption will also result in several problems challenging the
magnetar model (Mereghetti 2008; Tong \& Xu 2011).
\begin{enumerate}
  \item The spin down time scale of a newly born magnetar will be less than the shock breakout time
  due to the presence of a strong dipole magnetic field. This will cause the supernovae associated
  with magnetars more energetic than canonical supernovae (Duncan \& Thompson 1992). However, observations
  of supernova remnants associated with AXPs and SGRs show that the corresponding supernova energies are
  of canonical value (Vink \& Kuiper 2006). This failed prediction of the magnetar model may be circumvented
  if the initial rotational energy of magnetars are carried away in non-electromagnetic form, e.g., gravitational
  waves (Dall'Osso et al. 2009). However, in Dall'Osso et al. (2009), a relatively low dipole magnetic field is
  also required ($B_{\rm dip} \le 10^{14} \, \rm G$). If magnetars have a different braking mecahnism and 
  consequently their dipole magnetic field is much lower, this may explain their supernova energy problem.

  \item If AXPs and SGRs are neutron stars with strong dipole field, then although they rotate rather slowly
  (periods: $2$--$12$ seconds), they will also accelerate particles to very high energy. In the outer magnetosphere,
  these particles will emit high-energy gamma-rays which are detectable by $\it Fermi$-LAT (Cheng \& Zhang 2001).
  This may be viewed as an independent measurement of strong dipole magnetic field, i.e., through unipolar induction effect.
  However, $\it Fermi$-LAT observations of all AXPs and SGRs show no significant detection (Sasmaz Mus \& Gogus 2010;
  Abdo et al. 2010). Therefore, there are conflicts between the out gap model in the case of magnetars and
  {\it Fermi}-LAT observations (Tong et al. 2010a, 2011). It is possible that magnetars have a different  braking
  mechanism and their dipole magnetic field is not so strong.

  \item In the traditional picture of the magnetar model, magnetars are young neutron stars with both strong
  dipole field and strong multipole field (Thompson \& Duncan 1995, 1996; Thompson et al. 2002). The observation of the low magnetic
  field soft gamma-ray repeater SGR 0418+5729 has challenged the traditional magnetar prescription (Rea et al. 2010).
  This source tells us that magnetar-like activities (an anomalous X-ray luminosity or SGR-type bursts) do not require
  a strong dipole magnetic field. The timing of SGR Swift J1822.3--1606 further strengthens this point (Rea et al. 2012a).
  The originally required strong dipole magnetic field in most AXPs and SGRs mainly provides
  the braking torque. It is possible that not only SGR 0418+5729 but also many other AXPs and SGRs do not have 
  a strong dipole magnetic field if they have a different braking mechanism.

  \item There are high magnetic field rotation-powered pulsars (HBPSRs) along with magnetars (Ng \& Kaspi 2011).
  Although they are close to each other on the $P$--$\dot{P}$ diagram, they show very different timing behavior.
  The timing behaviors of HBPSRs are similar to that of normal pulsars (Ng \& Kaspi 2011). Therefore,
  it is reasonable that they have the same braking mechanism as that of normal pulsars.
  However, magnetars are very noisy (Gavriil \& Kaspi 2002; Woods et al. 2002;
  Archibald et al. 2008), and the period derivatives of magnetars can vary significantly (up to a factor of 10,
  Gavriil \& Kaspi 2004; Camilo et al. 2007; Woods et al. 2007).
  Therefore, it is possible that magnetars have a different braking mechanism, e.g. wind braking.
  The variation of wind luminosity will cause the variation of their period derivatives.

  \end{enumerate}

All these issues are related to the dipole magnetic field and the braking mechanism of magnetars.
A different braking mechanism of magnetars may help to solve these problems.
Electrodynamics of magnetars show that they may have globally twisted magnetospheres
(Thompson et al. 2002; Beloborodov \& Thompson 2007). The twisted magnetosphere will enhance their spin-down
torque and also modify their persistent emissions. Changes in their global magnetospheric structure will
result in changes in their spin-down rate and persistent flux (Beloborodov 2009).
In the case of wind braking, the large scale dipole field is
unchanged. It is the change of the particle wind luminosity that causes change of the spin-down rate.
Since both their persistent emissions and the particle wind are magnetism-powered,
it is natural that their spin-down behavior and persistent emissions are correlated.
Based on previous researches (Harding et al. 1999; Thompson et al. 2000), we explore the wind braking of magnetars
in more detail and apply it to all AXPs and SGRs. A comparison with up-to-date observations is also given.

Observations supporting the existence of a particle wind in magnetars are given in Section 2. 
Rotational energy loss rate due to a particle wind is calculated in Section 3. 
Several aspects of wind braking of magnetars are given in Section 4.
Discussions and conclusions are presented in Section 5 and 6, respectively.

\section{Existence of a particle wind}

\subsection{Qualitative description of wind braking of magnetars}

In the magnetic dipole braking scenario of normal pulsars, the star's rotational energy is carried away
by magnetic dipole radiation plus a rotation-powered particle wind (Michel 1969;
Xu \& Qiao 2001; Spitkovsky 2006). The rotational energy loss rate is quantitatively similar to
the magnetic dipole radiation in vacuum (Xu \& Qiao 2001; Spitkovsky 2006). The surface dipole magnetic field is almost the same
as that of magnetic dipole braking in vacuum. A particle wind mainly causes 
higher order modifications of pulsar timing, e.g. braking index (Michel 1969;
Manchester et al. 1985; Xu \& Qiao 2001; Contopoulos \& Spitkovsky 2006; Wang et al. 2012)
and timing noise (Lyne et al. 2010; Liu et al. 2011). Around young neutron stars, we may see a
rotation-powered pulsar wind nebula (Gaensler \& Slane 2006).

In the case of magnetars, the star's persistent X-ray luminosity is much higher than its rotational
energy loss rate. Since the persistent X-ray luminosity is from magnetic field decay, it is possible
that a particle flow (i.e., a magnetism-powered particle wind) is also produced during the decay of the star's magnetic field.
The luminosity of this particle wind\footnote{By saying ``particle wind'', 
we always mean a mixture of relativistic (or mildly relativistic) particles and electromagnetic waves.
In Section 3, we will point out the difference of particle luminosity and wind luminosity. 
At present and for general discussions, we will simply use the term ``particle wind''.} 
can be as high as the star's persistent X-ray luminosity,
therefore it can also be much higher than the star's rotational energy loss rate (Duncan 2000; Section 2.3 below). This particle wind will
``comb out'' the magnetic field lines in the closed field line regions (Harding et al. 1999).
The net result is an enhanced rotational energy loss rate for a given dipole magnetic field (Harding et al. 1999;
Thompson et al. 2000; Section 3 below). In this ``wind aided'' spin down scenario, the corresponding dipole magnetic field
will be much lower than the magnetic dipole braking case (Harding et al. 1999; Section 4 below).
Wind braking of magnetars will also help us to explain recent observations challenging the the
traditional magnetar model (Section 4 and 5 below).

Below, we assume that the star's dipole magnetic field is constant during its life time. It is the evolution
and variation of particle wind luminosity that cause the evolution and variation of AXPs/SGRs' timing properties.
For example, a short term variation of particle wind will cause a variation of the star's period derivative and also
contribute to its timing noise.

\subsection{Observational clues for the existence of a particle wind}

The existence of a (rotation-powered) particle wind in normal pulsars is
well established. The observations of intermittent pulsars give direct support for
the existence a particle wind (Kramer et al. 2006; Camilo et al. 2012). However, the existence of a
magnetism-powered particle wind in magnetars is still unknown. Below
we will give several observations of AXPs and SGRs, which may provide some hints for
the existence of a particle wind.
\begin{enumerate}
  \item The AXP 1E 2259+586 experiences an enhanced period of spin down during outburst
  (Kaspi et al. 2003). Variations of period derivative are also seen in AXP 1E 1048.1--5937
  (Gavriil \& Kaspi 2004), AXP XTE J1810--197 (Camilo et al. 2007), SGR 1806--20 (Woods et al. 2007)
  and AXP 1E 1547.0--5408 (Camilo et al. 2008) etc.
  A decreasing period derivative is also observed in the radio-loud magnetar, accompanying by a
  decaying X-ray luminosity and radio luminosity (Levin et al. 2012; Anderson et al. 2012).
  This may be due to a decaying particle wind during outbursts.

  In the absence of a strong particle wind, an untwisting magnetosphere of a magnetar
  may explain the decreasing period derivative (Beloborodov 2009). However the dipole field is of large scale compared with that of multipole
  field, a varying dipole field (especially short time scale variations) is hard to accomplish (Camilo et al. 2007; Levin et al. 2012).
  In the case of
  wind braking of magnetars, the global dipole field is unchanged.
  Since the particle wind may be the consequences of small amplitude seismic activities (Thompson \& Duncan 1996),
  it can vary dramatically even on short time scales.
  A varying particle wind  will cause a varying period derivative. The long term decay of particle wind luminosity
  during outburst can account for the decreasing period derivative (e.g., AXP XTE J1810--197, Camilo et al. 2007;
  the radio-loud magnetar, Levin et al. 2012).
  During an outburst, we should expect
  the wind luminosity first increases then decreases. This will cause the period derivative first increases
  then decreases, which may be the case of SGR 1806-20 (Woods et al. 2007) and AXP 1E 1547.0-5408 (before outburst, Camilo et al. 2008).

  \item AXPs and SGRs have a higher level of timing noise than normal pulsars (Gavriil \& Kaspi 2002;
  Woods et al. 2002; Archibald et al. 2008). The timing noise may be correlated with period derivatives.
  The timing noise of normal pulsars may be the result of a varying (rotation-powered) particle wind (Lyne et al. 2010;
  Liu et al. 2011). Then it is possible that AXPs and SGRs are also braked down by a particle wind.
  Since AXPs and SGRs are magnetism-powered, the particle wind may also from magnetic field decay, i.e., a magnetism-powered
  particle wind. This magnetism-powered particle wind may vary significantly with time, similar to the magnetar's persistent
  X-ray luminosity. Then it may cause a higher level of timing noise in magnetars than that in 
  normal pulsars and HBPSRs.

  \item If AXPs and SGRs harbor a strong enough particle wind (either rotation-powered or magnetism-powered),
  then we should see a pulsar wind nebula around the putative star. If the the particle wind is magnetism-powered
  , the same as that of their persistent X-ray luminosities, then we should see some correlation between the nebula luminosity
  and the stellar luminosity. A possible extended emission is found around AXP 1E 1547.0--5408
  (Vind \& Bamba 2009). The luminosity of the extended emission is correlated with the star's luminosity
  (Olausen et al. 2011). Therefore, the extended emission around AXP 1E 1547.0--5408 may be a magnetism-powered pulsar
  wind nebula instead of a duct scattering halo. If this is confirmed in the future, 
  it will be a strong evidence for the existence a magnetism-powered particle wind in magnetars.

  A magnetism-powered pulsar wind nebula may also accelerate particles to very high energy and radiate high
  energy photons. An extended emission and a TeV source are both seen in the case of SGR Swift J1834.9--0846 (Kargaltsev et al. 2012).
  If the extended emission is found to be a pulsar wind nebula and the associated with the TeV source is confirmed,
  then it is also likely to be a magnetism-powered pulsar wind nebula\footnote{After we put this paper on the arXiv (1205.1626),
  Younes et al. (arXiv:1206.3330) propose that the extended emission of SGR Swift J1834.9--0846 
  may be a magnetism-powered pulsar wind nebula (since it has a high conversion efficiency).
  This observation is consistent our analysis here.}.
  A candidate pulsar wind nebula which may contain magnetic energy contribution
  is seen around RRAT J1819--1458 (Rea et al. 2009).
\end{enumerate}

In summary, there are many uncertainties and ambiguities if we attribute the above observations to a particle wind in magnetars.
However, we do not know  whether AXPs and SGRs have a (magnetism-powered) particle wind or not.
The possibility of such a particle wind can not be ruled out by present observations either.
A magnetism-powered particle wind in magnetars is helpful to our understanding of the different observational aspects stated above.
Therefore, the above observational facts may give us some clues for the existence of a particle wind in magnetars.
Whether a particle wind really exists or not can be tested by future studies.

\subsection{Estimation of wind luminosity}

In the magnetar model, the bursts and outbursts are related with the magnetar's seismic
activities (Thompson \& Duncan 1995, 1996). If the observable
bursts are associated with large amplitude seismic activities, then the low amplitude seismic
activities may mainly result in a particle wind (Thompson \& Duncan 1996). According to Thompson \& Duncan
(1996, eq.(71) there), the particle wind luminosity is
\begin{equation}
L_{\rm p} \simeq 2 \times 10^{35} \left ( \frac{B_{\rm c}}{10^{15} \,\rm G} \right )^2
\left ( \frac{t}{10^{4} \,\rm yr} \right)^{-1} \left ( \frac{\Delta R_{\rm c}}{1 \,\rm km} \right ) \, \rm erg \, s^{-1},
\end{equation}
where $B_{\rm c}$ is crustal field strength, $t$ is the star's age, and $\Delta R_{\rm c}$ is the
crustal thickness. The above equation is only valid for crustal field strength less than $6\times 10^{15} \, \rm G$,
above which the crust may undergo plastic deformations.

The persistent X-ray luminosity of AXPs and SGRs are from magnetic field decay, e.g., internal heating
(Thompson \& Duncan 1996) or magnetospheric current heating (Thomspon et a. 2002;
Beloborodov \& Thompson 2007). A particle wind may also be produced during this process. Since
the particle wind and the persistent X-ray luminosity are from the same energy reservoir,
a natural estimation of the particle wind luminosity is (Duncan 2000)
\begin{equation}
L_{\rm p} \sim L_{\rm x} \sim 10^{35} \,\rm erg \, s^{-1},
\end{equation}
which is valid for most AXPs and SGRs. For the transient magnetars, they have a lower
quiescent X-ray luminosity. Their particle wind luminosity may also be correspondingly lower.

In the wind braking scenario, magnetars are neutron stars with strong multipole fields. The strong
twisted mangetic field in the vicinity of magnetars will accelerate particles to very high energy.
Thus, a corona of high energy particle will be formed (Beloborodov \& Thompson 2007). The footprint
of magnetic field lines are anchored to the stellar crust. In the presence of frequent low amplitude
seismic activities, the corona of magnetars will be disturbed continuously.
The excitation of such a particle wind in magnetars may be due to their seismic activities, especially
small amplitude seismic activities (Thompson \& Duncan 1996; Thompson \& Duncan 2001; Timokhin et al. 2008).
The particles in  the magnetar magnetosphere can flow out in two ways. (1) During bursts and giant flares.
This burst component of particle wind has its duty cycles (Thompson \& Duncan 1995; see numerical simulations
of Parfrey et al. 2012; Section 4.5 below). 
(2) During persistent state. The long term average of many small amplitude seismic activities
may result in a persistent particle outflow of magnetars (Thompson \& Duncan 1996; Duncan 2000). 
We will mainly focus on the persistent component of particle wind.

In conclusion, we have already
some observational clues for the existence of a particle wind in magnetars. Their luminosities can also
be estimated, although the underlying mechanism is still lacking. Since both the magnetar's persistent X-ray luminosity
and the particle wind are from magnetic field decay, 
the particle wind luminosity may be as high as their persistent X-ray luminosities.
Therefore, the particle wind luminosity in magnetars can be much higher than their rotational energy loss rate. 
The existence of such a strong particle wind will modify the spin-down behavior of magnetars qualitatively.

\section{Rotational energy loss rate due to a particle wind}

\subsection{Description of the global magnetospheric structure}

The magnetospheres of pulsars and magnetars contain regions of open and closed magnetic field. 
The closed field lines extend to the light cylinder radius in the case of normal pulsars (Contopoulos \& Spitkovksy 2006). 
In the case of magnetars, the closed field line region may be smaller. 
In the presence of a strong particle wind, the natural radial extension of closed field line regions is the radius 
where the kinetic energy density of particle wind equals the magnetic energy density (Harding et al. 1999; Thompson et al. 2000). 
Particle flows in the closed field line regions belong to the domain of closed field line region 
electrodynamics of magnetars (Thomspon et al. 2002; Beloborodov \& Thompson 2007; Tong et al. 2010b). 
Particle flow collimated around the polar cap may dominate the 
spin down of the central star. The opening angle of the polar cap region is determined by the coupling between 
the magnetar crust and its magnetosphere. The total particle luminosity $L_{\rm p}$ is determined by 
the decay of magnetic field energy. Only a fraction of this particle wind can flow out to infinity and contribute to 
the spin down of the magnetar. The escaping particle luminosity is denoted as $L_{\rm w}$, i.e., wind luminosity. 
Then it is natural that $L_{\rm w} \le L_{\rm p}$. For a given particle luminosity, the maximum braking case
is accomplished when the wind luminosity equals the total particle luminosity.

\subsection{The simplest case: $L_{\rm w} =L_{\rm p}$}

For a neutron star with angular velocity $\Omega=2\pi/P$
($P$ is rotation period), its light cylinder radius $R_{\rm{lc}}$ is (the
radius where the rotational velocity equals the speed of light)
\begin{equation}
 R_{\rm{lc}} = \frac{c}{\Omega} =\frac{Pc}{2\pi} =4.8\times 10^{10} \left( \frac{P}{10\,\rm s} \right) \,\rm cm,
\end{equation}
where $c$ is the speed of light. In the case of magnetars, with the aid of a particle
wind, the magnetic field lines are combed out at a radius $r_{\rm{open}}$ (where the particle energy density
equals the magnetic energy density, Harding et al. 1999)
\begin{equation}\label{r_open}
 r_{\rm{open}} = r_0 \left( \frac{B_0^2 r_0^2 c}{2 L_{\rm w}} \right)^{1/4}
 = r_0 \left( \frac{B_0^2 r_0^2 c}{2 L_{\rm p}} \right)^{1/4}
 =4.1\times 10^9 b_0^{1/2} L_{\rm p,35}^{-1/4} \,\rm cm,
\end{equation}
where $r_0=10^6 \,\rm cm$ is neutron star radius, $B_0=b_0 \times B_{\rm QED}$ is dipole magnetic field at the magnetic pole,
and $L_{\rm w} = L_{\rm p} =L_{\rm p,35}\times 10^{35} \, \rm erg\, s^{-1}$ is the particle wind luminosity 
(assuming\footnote{This means that there is only a small particle flow in the closed field line regions.} $L_{\rm w} =L_{\rm p}$, 
and assuming the escaping particle wind becomes near isotropic at $r_{\rm open}$). 
The polar cap radius now is
\begin{equation}\label{Rpc_wind_braking}
 R_{\rm{pc}} = r_0 (r_0/r_{\rm{open}})^{1/2} = 1.6\times 10^4 b_0^{-1/4} L_{\rm p,35}^{1/8} \,\rm cm.
\end{equation}
The corresponding polar cap opening angle is 
\begin{equation}\label{thetaopen}
\theta_{\rm open}^2= r_0/r_{\rm open}=2.4\times 10^{-4} \,b_0^{-1/2} \, L_{\rm p,35}^{1/4}.
\end{equation}
Typically, $\theta_{\rm open} = 1.6\times10^{-2} \, b_0^{-1/4} \, L_{\rm p,35}^{1/8}$. 
The polar cap opening angle $\theta_{\rm open}$ depends on the wind luminosity $L_{\rm w}$.

This forms the basic structure of a wind-loaded magnetosphere.
The star may form a current circuit in the open field line regions.
The rotational energy loss rate
due to this particle wind is (Harding et al. 1999)
\begin{equation}\label{Edot_Harding}
 \dot{E}_{\rm{w}} =\frac{B_0^2 r_0^6 \Omega^4}{3 c^3} \left( \frac{R_{\rm{lc}}}{r_{\rm{open}}} \right)^2.
\end{equation}
For the traditional magnetic dipole braking,
the corresponding rotational energy loss rate is\footnote{
Note that eq.(\ref{Edot_D}) is for orthogonal rotators.
While in the wind braking case, for simplicity, we are considering an aligned rotator
(Harding et al. 1999).
Therefore, eq. (\ref{Edot_D}) should be taken as definition rather than derivation.}
(Shaprio \& Teukolsky 1983)
\begin{equation}\label{Edot_D}
 \dot{E}_{\rm{d}} =\frac{B_0^2 r_0^6 \Omega^4}{6 c^3}.
\end{equation}
Therefore, eq.(\ref{Edot_Harding}) can be rewritten as
\begin{equation}
 \dot{E}_{\rm{w}} =\frac{2}{\sqrt{3}} \dot{E}_{\rm{d}} \left( \frac{L_{\rm p}}{\dot{E}_{\rm{d}}} \right)^{1/2}.
\end{equation}

A second way to calculate the rotational energy loss rate due to a particle wind is provided by
Thompson et al. (2000)\footnote{$r_{\rm open}$ in Harding et al. (1999) is equivalent to the
Alfv\'{e}n radius $R_{\rm A}$ in Thompson et al. (2000), except for a difference of constant factor $2^{1/4}$.
Hereafter, $r_{\rm open}$ is employed.}. The outflowing particles will corotate with the star up
to the radius $r_{\rm open}$. For relativistic (also mildly relativistic) particles, the rotational
energy carried away by this particle wind is (Thompson et al. 2000)
\begin{equation}
 \dot{E}_{\rm w} =\frac23 \frac{L_{\rm p}}{c^2} \Omega^2 r_{\rm open}^2
=\frac{2}{\sqrt{3}} \dot{E}_{\rm{d}} \left( \frac{L_{\rm p}}{\dot{E}_{\rm{d}}} \right)^{1/2}.
\end{equation}

A third way to calculate the rotational energy loss rate due to a particle wind can be done
in analogy with that of Xu \& Qiao (2001). The electric current in the two polar caps will
carry away the rotational energy of the star in the presence of an acceleration potential.
This acceleration potential is due to unipolar induction.
Assuming maximum acceleration potential, the rotational energy loss rate is
\begin{equation}
 \dot{E}_{\rm{w}} =2I_{\rm pc} \Phi_{\rm max}
=\frac{3}{\sqrt{3}} \dot{E}_{\rm{d}} \left( \frac{L_{\rm p}}{\dot{E}_{\rm{d}}} \right)^{1/2},
\end{equation}
where $I_{\rm pc} =\pi R_{\rm pc}^2 \rho_{\rm GJ} c$ is the polar cap current (for one polar cap), $\rho_{\rm GJ}$
is the Goldreich-Julian density, and $\Phi_{\rm max}$ is the maximum acceleration potential
due to unipolar induction (Ruderman \& Sutherland 1975)
\begin{equation}
 \Phi_{\rm max} =\frac{B_0 r_0^2 \Omega }{2c} \left( \frac{R_{\rm pc}}{r_0} \right)^2.
\end{equation}

Therefore, irrespective of the details of particle wind, accurate within a factor of two,
the rotational energy loss rate due to a particle wind can be written as (Harding et al. 1999)
\begin{equation}\label{Edot_w}
 \dot{E}_{\rm{w}} =\dot{E}_{\rm{d}} \left( \frac{L_{\rm w}}{\dot{E}_{\rm{d}}} \right)^{1/2}
 =\dot{E}_{\rm{d}} \left( \frac{L_{\rm p}}{\dot{E}_{\rm{d}}} \right)^{1/2}
\end{equation}
The second identity is obtained by assuming $L_{\rm w} =L_{\rm p}$. 

From equation (\ref{Edot_w}), we see that
\begin{enumerate}
  \item For a rotation-powered particle wind $L_{\rm p} \sim - \dot{E}_{\rm rot}$,
  $\dot{E}_{\rm w} \sim \dot{E}_{\rm d} \sim -\dot{E}_{\rm rot}$, wind braking is quantitatively similar to the
  case of magnetic dipole braking in vacuum. The effects of particle wind will mainly cause higher order
  modifications, e.g., a different braking index etc. This is the case of normal pulsars.
  
  \item For magnetars, there may be a magnetism-powered particle wind $L_{\rm p} \gg - \dot{E}_{\rm rot}$.
  Wind braking of magnetars will result in $\dot{E}_{\rm w} = -\dot{E}_{\rm rot} \gg \dot{E}_{\rm d}$.
  Therefore, the magnetic dipole braking is enhanced due to the presence of a
  magnetism-powered particle wind (Harding et al. 1999). This will cause a strong reduction of magnetar's 
  dipole field. Meanwhile, high order effects will also exists, e.g., a different braking index, 
  larger timing noise, a magnetism-powered pulsar wind nebula, etc.
\end{enumerate}

\subsection{Detailed considerations of wind luminosity}
\label{section_Lw}

The above simplest case assumes the escaping wind luminosity is equal to the total particle luminosity.
From equation (\ref{thetaopen}), the polar opening angle depends on the escaping wind luminosity. 
This means that the polar cap opening angle (at the star surface) is affected by the physics happening at $r_{\rm open}$. 
It is not known how this is accomplished. Alternatively, the polar cap opening angle of the particle wind may be an 
independent parameter. The total particle luminosity may involve a particular angular distribution. 
This angular distribution may result from coupling between the magnetar crust and its magnetosphere.
The typical time scale of this coupling may be estimated from quasi-periodic
oscillations in magnetars (Timokhin et al. 2008; Watts 2011). The fundamental frequency is about
$\nu\sim 20 \,\rm Hz$. The length scale of coupling between the neutron star and its magnetosphere
is
\begin{equation}
r_{\rm max} \sim \frac{c}{3\nu} \sim 5\times 10^8 \left( \frac{20\,\rm Hz}{\nu} \right) \,\rm cm.
\end{equation}
The corresponding polar cap opening angle is
\begin{equation}
\theta_{\rm s}^2 = \frac{r_0}{r_{\rm max}} \sim 2\times 10^{-3} \left( \frac{\nu}{20\,\rm Hz} \right).
\end{equation}
Typically, $\theta_{\rm s} \sim 0.05 \left( \frac{\nu}{20\,\rm Hz} \right)^{1/2}$. The particles
will mainly flow through the polar cap area with opening angle $\theta_{\rm s}$. In the following
calculations, we will take $\theta_{\rm s}$ as the fundamental input parameter. $r_{\rm max}$ etc will be
functions of $\theta_{\rm s}$.

The particles from the two polar cap regions can flow out to radius larger than $r_{\rm max}$.
Considering the presence of strong magnetic field, a significant amount of the outflowing particles
may be trapped in the closed field line regions in the magnetosphere\footnote{These trapped particles
may contribute to the persistent X-ray emissions of magnetars.}. Only a fraction of them can
flow out to infinity and therefore carry away the star's rotational energy. The Alfv\'{e}n radius
characterize the effect of magnetic field quantitatively. We denote it as $r_{\rm open}$ in accordance
with equation (\ref{r_open}). In the present case, 
it is also defined as the radius when the particle energy density equals the magnetic energy density
\begin{equation}
\gamma \rho(r) c^2 \sim \frac{B(r)^2}{8\pi},
\end{equation}
where $\gamma$ and $\rho(r)$ are the Lorentz factor and mass density, respectively.
When particles move along magnetic field lines, their kinetic energy is conserved
(not considering radiation losses). The mass density may scale with the local Goldreich-Julian
charge density
$\rho(r) \propto \rho_{\rm GJ} \propto 1/r^3$. Therefore
\begin{equation}
\gamma \rho_{\rm s} c^2 \left( \frac{r_0}{r_{\rm open}} \right)^3
\sim \frac{B_0^2}{8\pi} \left( \frac{r_0}{r_{\rm open}} \right)^6,
\end{equation}
where $\rho_{\rm s}$ is the mass density at the star surface.
According to the definition of particle luminosity and assuming uniform distribution across
the polar cap region
\begin{equation}
L_{\rm p} = 2\pi (r_0 \theta_{\rm s})^2 \gamma \rho_{\rm s} c^2 \,c,
\end{equation}
then $r_{\rm open}$ is
\begin{eqnarray}
r_{\rm open} &= &r_0 \left( \frac{B_0^2}{8\pi} \frac{2\pi (r_0 \theta_{\rm s})^2 \,c}{L_{\rm p}} \right)^{1/3}\\
&=& 7\times 10^9 \,b_0^{2/3} L_{\rm p,35}^{-1/3} \left( \theta_{\rm s}/0.05 \right)^{2/3} \,\rm cm.
\end{eqnarray}

Only the escaping wind particles can carry away the star's rotational energy. From the definition of wind luminosity,
$L_{\rm w} \propto \theta_{\rm open}^2 \propto 1/r_{\rm open}$. At the same time, the total particle luminosity
is $L_{\rm p} \propto \theta_{\rm s}^2 \propto 1/r_{\rm max}$. The wind luminosity is related to the total
particle luminosity
\begin{equation}
L_{\rm w} = L_{\rm p} \frac{r_{\rm max}}{r_{\rm open}}.
\end{equation}
Taken the polar cap opening angle as the fundamental parameter, $r_{\rm max}$ will be
$r_{\rm max} =r_0/\theta_{\rm s}^2$. Therefore, the wind luminosity is
\begin{equation}\label{wind luminosity}
L_{\rm w} = 6\times 10^{33} \, b_0^{-2/3} L_{\rm p, 35}^{4/3} (\theta_{\rm s}/0.05)^{-8/3} \,\rm erg \,s^{-1}.
\end{equation}
The wind luminosity depends strongly on the polar cap opening angle, i.e., how the neutron star couples with
the magnetosphere. In the present case, the wind luminosity is a fraction of the total particle
luminosity. Then, it must be that $L_{\rm w}\leq L_{\rm p}$. In terms of $r_{\rm max}$ and $r_{\rm open}$,
it must be that $r_{\rm max} \leq r_{\rm open}$. 

The calculation of rotational energy loss rate is the same as the previous section. 
From equation (\ref{Edot_w}), the rotational energy loss rate due to a particle wind in the present case is
\begin{equation}
\dot{E}_{\rm{w}} =\dot{E}_{\rm{d}} \left( \frac{L_{\rm w}}{\dot{E}_{\rm{d}}} \right)^{1/2},
\end{equation}
where $L_{\rm w}$ is determined by equation (\ref{wind luminosity}). 
The neutron star's dipole magnetic field is obtained by equaling $-\dot{E}_{\rm rot} = \dot{E}_{\rm w}$,
\begin{eqnarray}
\nonumber
B_0 &=& 3.3 \times 10^{32} \left( \frac{\dot{P}}{P} \right)^{3/2}
L_{\rm p,35}^{-1} (\theta_{\rm s}/0.05)^2 \rm \, G\\
&=& 3.3 \times 10^{14} \left( \frac{\dot{P}/10^{-11}}{P/10\,\rm s} \right)^{3/2}
L_{\rm p,35}^{-1} (\theta_{\rm s}/0.05)^2 \rm \, G.
\end{eqnarray}
The dipole magnetic field is determined by four parameters: the period and its derivative,
the total particle luminosity, and the polar cap opening angle. If the polar cap opening angle
is three times smaller, the dipole magnetic field will be ten times lower.

In conclusion, considering detailed modeling of wind luminosity, the rotational energy
loss rate is reduced compared with the simplest case. The model parameter space is larger with 
the addition of another variable $\theta_{\rm s}$. 
There are parameter space that the corresponding
dipole magnetic field is only slightly lower than the magnetic dipole braking case. At the same time, there
are also some parameter space that the dipole magnetic field is much lower than the magnetic dipole braking
case. The following calculations in Section 4 are mainly done in the simplest case. This corresponds to maximum
braking for a given particle luminosity. In this way, we want to demonstrate to which extent can wind braking
of magnetars help to explain the current observations. For the calculations in Section 4.2, 4.4, and 4.5,
the conclusions are unaffected by different assumptions. For the calculations in Section 4.1 and 4.3,
the results may only change quantitatively.

\section{Wind braking of magnetars}

Wind braking of magnetars had been considered previously by Marsden et al. (1999, for the case of SGR 1900+14),
Harding et al. (1999, for the case of SGR 1806--20), Thompson et al. (2000, for the case of SGR 1900+14).
They mainly talked about wind braking during outbursts, although some of the formulae of
the long term wind-aided spin down are also given by Thompson et al. (2000, Section 4.1 there).
We explore the wind braking in more details and apply it to all magnetars.
A comparison with up-to-date observations is also presented.

\subsection{Dipole magnetic field}

From eq. (\ref{r_open}) and eq. (\ref{Edot_w}), the presence of particle wind amplifies the magnetic dipole braking
rotational energy loss rate. Therefore, wind braking is valid only when
\begin{equation}\label{LpEdotD}
 L_{\rm p} \geq \dot{E}_{\rm{d}}.
\end{equation}
Equaling the rotational energy loss rate $-\dot{E}_{\rm rot}$ ($=-I \Omega \dot{\Omega}$) and eq. (\ref{Edot_w}), we get
\begin{equation}\label{validity}
 -\dot{E}_{\rm rot} =\dot{E}_{\rm{w}} \leq L_{\rm p}.
\end{equation}
Wind braking of magnetars is valid only when the wind luminosity is greater than the star's rotational
energy loss rate. Equation (\ref{validity}) can be rewritten as
\begin{equation}
 -\dot{E}_{\rm rot} =\dot{E}_{\rm w} \geq \dot{E}_{\rm{d}}.
\end{equation}
The characteristic magnetic field obtained by assuming magnetic dipole braking
is only the upper limit of the star's true dipole magnetic field.

Assuming magnetic dipole braking
\begin{equation}
 -\dot{E}_{\rm rot} =\dot{E}_{\rm{d}} =\frac{B_0^2 r_0^6 \Omega^4}{6 c^3},
\end{equation}
the dipole magnetic field (at the magnetic pole) is
\begin{equation}
 B_0 =6.4\times 10^{19} \sqrt{P \dot{P}} \,{\rm G}
= 6.4\times 10^{14} \left( \frac{P}{10 \,\rm{s}} \frac{\dot{P}}{10^{-11}}\right)^{1/2} \,{\rm G}.
\end{equation}
It is two times larger than usually reported since the polar magnetic field is two times larger
than the equatorial magnetic field (eq. (5.17) in Lyne \& Graham-Smith 2012 and corresponding discussions).
However, the above magnetic dipole braking is originally designed for rotation-powered pulsars.
Magnetars may be wind braking instead of magnetic dipole braking, as discussed above.
In the case of wind braking
\begin{equation}\label{Bdipw1}
 -\dot{E}_{\rm rot} =\dot{E}_{\rm{w}} =\dot{E}_{\rm{d}} \left( \frac{L_{\rm p}}{\dot{E}_{\rm{d}}} \right)^{1/2}. 
\end{equation}
The corresponding dipole magnetic field is
\begin{equation}\label{Bdipw2}
 B_0 =4.0\times 10^{25} \frac{\dot{P}}{P} \,L_{\rm p,35}^{-1/2} \,{\rm G}
=4.0\times 10^{13} \frac{\dot{P}/10^{-11}}{P/10\,\rm{s}} \,L_{\rm p,35}^{-1/2} \,{\rm G}.
\end{equation}
For typical AXPs and SGRs, the dipole magnetic field in the case of wind braking
is about ten times lower than that of magnetic dipole braking. Therefore, AXPs and
SGRs may be magnetars without strong dipole field. Only strong multipole field
($\sim 10^{14}-10^{15} \,\rm{G}$) is required to power their bursts, persistent
emissions, and braking.

At the time when Harding et al. wrote their wind braking
paper (Harding et al. 1999), they did not realize that there are two kinds of magnetic fields in
magnetars: dipole field and multipole field. When they saw that a strong dipole field is not
needed in the case of winding braking, Harding et al. said that ``the magnetar model must be
abandoned'' as the penalty of wind braking. With the presence of multipole field, AXPs and SGRs
can also show magnetar-like activites without a strong dipole field. This point is demonstrated clearly
by the observation of SGR 0418+5729 (Rea et al. 2010). The timing of SGR Swift J1822.3$-$1606
further strengthens this point (Rea et al. 2012a).

\begin{figure}[!htbp]
 \centering
\includegraphics{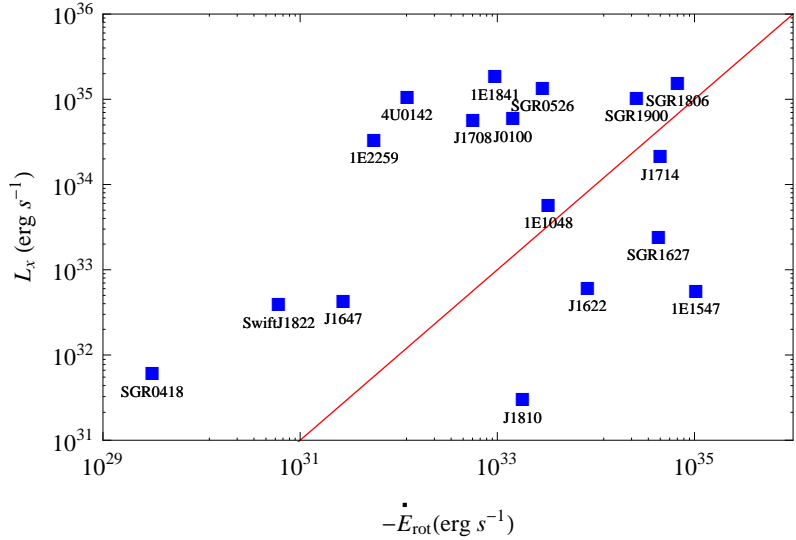}
\caption{Persistent X-ray luminosities of magnetars versus their spin down luminosities.
The solid line is $L_{\rm x}=-\dot{E}_{\rm rot}$. See table \ref{table_Bdip} and text for details.}
\label{fig_LxEdot}
\end{figure}

Table \ref{table_Bdip} summaries the observed parameters and deduced quantities for all AXPs and SGRs (17 in total),
which have period, period derivative, and persistent X-ray luminosity measured.
Figure \ref{fig_LxEdot} shows the magnetar persistent X-ray luminosity versus the star's
rotational energy loss rate. We employ the following two ways to the model the particle luminosity from magnetars.
\begin{enumerate}
  \item Considering that for all AXPs and SGRs, they must have a strong multipole field ($\sim 10^{14}-10^{15} \,\rm G$) in order
  to show magnetar-like activities. This is also true for low magnetic field SGRs (Rea et al. 2010, 2012a). Therefore, if the total
  field strength determines the particle luminosity, the particle luminosity will be more or less the same for all magnetars.
  In this case, we assume a particle luminosity $L_{\rm p}= 10^{35} \,\rm erg\, s^{-1}$ for all sources.
  From Figure \ref{fig_LxEdot}, we see that all AXPs and SGRs are braked by a particle wind except AXP 1E 1547.0--5408.
  For AXP 1E 1547.0--5408, the effect of a particle wind will mainly result in high order spin down behaviors, e.g. 
  a magnetism-powered particle wind surrounding the putative magnetar\footnote{For the case of AXP 1E 1547.0--5408,
  since it rotational energy loss rate is also relatively large, the surrounding pulsar wind nebula may be
  a mixture of rotation-powered and magnetism-powered particle wind.}.

  \item On the other hand, different sources may have a different evolution history.
  Irrespective of the detailed wind mechanism, the magnetar's particle luminosities may follow their persistent X-ray luminosities.
  In this case, we assume that the particle luminosities are the same as their persistent X-ray luminosities.
  From Figure \ref{fig_LxEdot}, except for the five sources with $L_{\rm x}<-\dot{E}_{\rm rot}$,
  the rest of AXPs and SGRs are all braked down by a particle wind\footnote{this may explain
  the ``fundamental plane'' of magnetar radio emission, see Section 4.2 below.}.
\end{enumerate}
At present, we do not know the detailed mechanism of magnetar wind.
The actual case may lie between these two extremes.

\begin{figure}[!htbp]
 \centering
\includegraphics{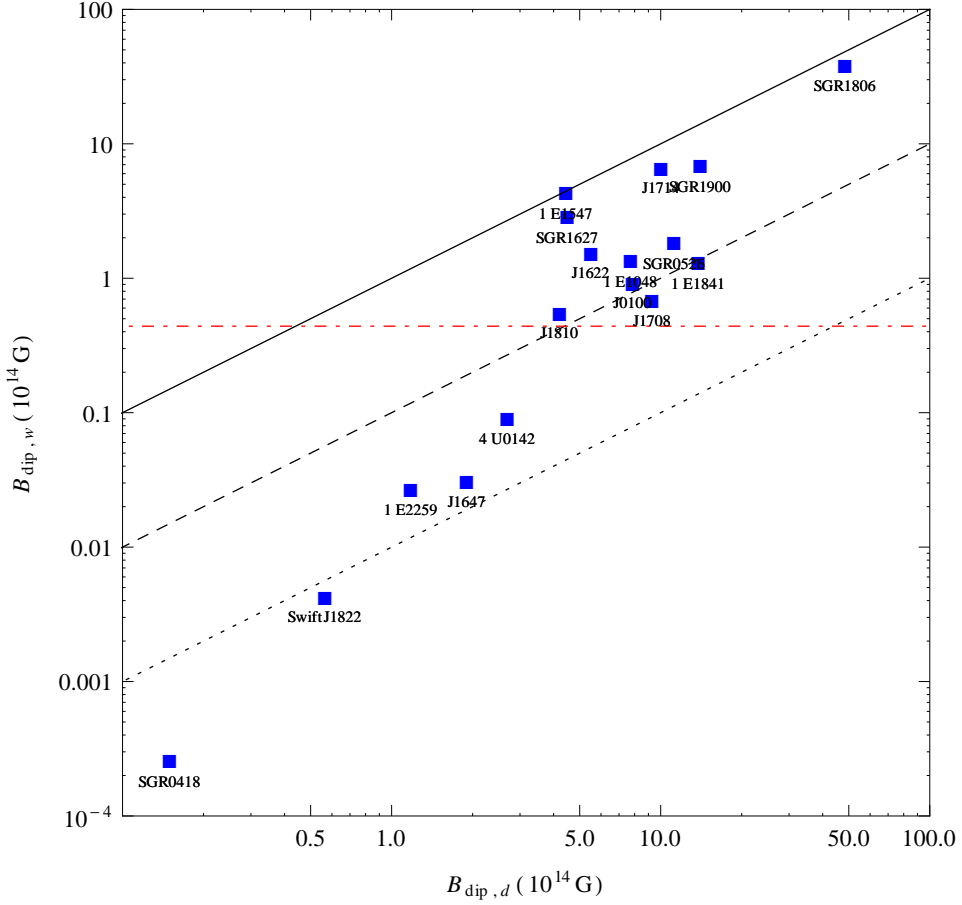}
\caption{Dipole magnetic field in the case of wind braking versus dipole magnetic field
in the case of magnetic dipole braking. A wind luminosity $L_{\rm p}= 10^{35} \,\rm erg\, s^{-1}$ is assumed for all sources.
The solid, dashed, and dotted lines are for
$B_{\rm dip,w}=B_{\rm dip,d},\ 0.1 B_{\rm dip,d},\ 0.01 B_{\rm dip,d}$, respectively.
The dot-dashed line marks the position of quantum critical magnetic field $B_{\rm QED}= 4.4\times 10^{13} \,\rm G$.}
\label{fig_Bdip1}
\end{figure}

\begin{figure}[!htbp]
 \centering
\includegraphics{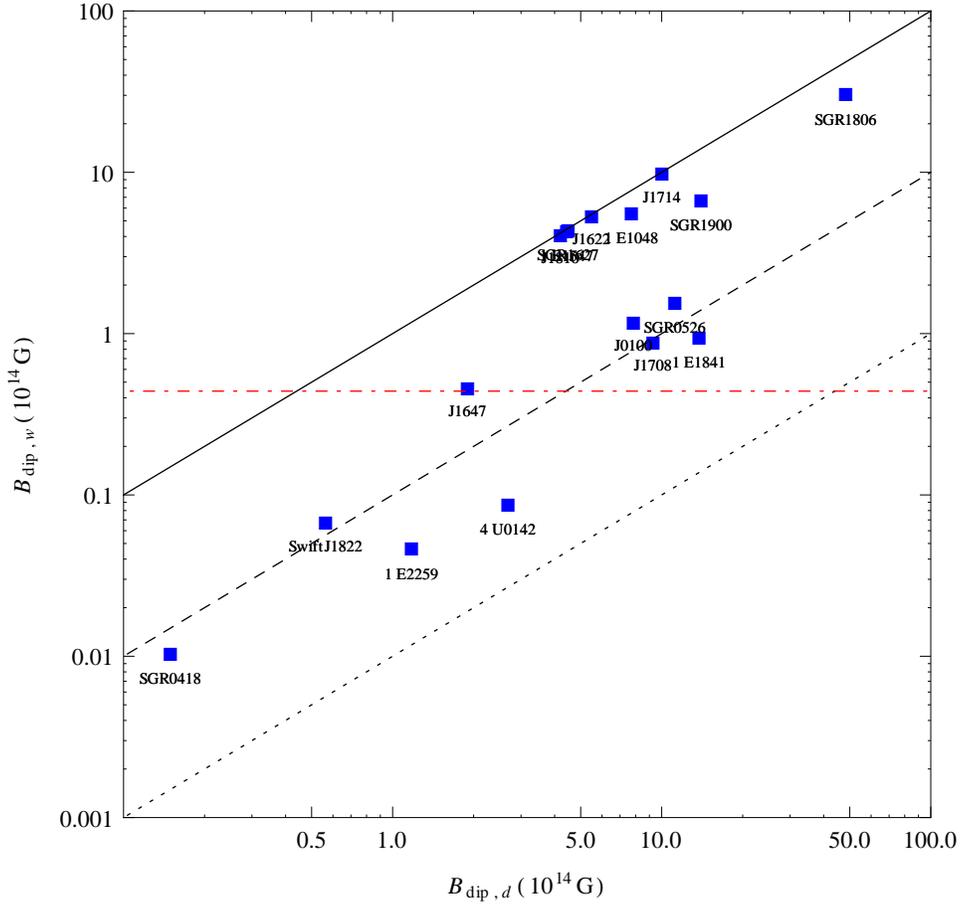}
\caption{The same as Figure \ref{fig_Bdip1}.
The wind luminosities are assumed to be the same as their persistent X-ray luminosities.
See text for details.}
\label{fig_Bdip2}
\end{figure}

Figure \ref{fig_Bdip1} and \ref{fig_Bdip2} show the dipole magnetic field in the case of wind braking versus the
dipole magnetic field in the case of magnetic dipole braking. From Figure \ref{fig_Bdip1} and \ref{fig_Bdip2},
we see that
\begin{enumerate}
 \item For most AXPs and SGRs, their dipole magnetic field by assuming wind braking
are ten time lower than that of magnetic dipole braking. This may help us to understand
why the magnetar supernova energies are of canonical value (Vink \& Kuiper 2006; Dall'Osso et al. 2009).

Numerical simulation of particle wind during magnetar bursts also suggests that the long term
averaged period derivative may be greatly amplified (Parfrey et al. 2012). The actual dipole
magnetic field may be significant lower than the magnetic dipole braking case. This is consistent with
our considerations here.

 \item The corresponding dipole magnetic field $B_{\rm dip,w}$ ranges from $10^{12} \,\rm G$
to $10^{15} \,\rm G$. A strong dipole magnetic field (Kouveliotou et al. 1998, $>B_{\rm QED}=4.4\times 10^{13} \,\rm G$)
is no longer a necessary input. In the wind braking scenario, magnetars are neutrons with strong multipole field.
For most sources, their dipole field may or may not be as strong as their multipole field.

 \item For several sources, their $B_{\rm dip,w}$ are in the range $10^{13} \,\rm{G}-10^{14} \,\rm G$.
This is similar to that that of X-ray dim isolated neutron stars (Kaplan \& van Kerkwijk 2011; Tong et al. 2010b). 
Therefore, when the magnetar
activities of these sources calm down, they will become X-ray dim isolated neutron stars naturally.

 \item There are now more low magnetic field magnetars,
with $B_{\rm dip,w} <4.4 \times 10^{13} \,\rm G$. Therefore, in the case of wind braking,
SGR 0418+5729 (Rea et al. 2010) is not such peculiar as before.
\end{enumerate}

Furthermore, from eq.(\ref{Bdipw1}) and eq. (\ref{Bdipw2}) we see that for a given magnetar
\begin{enumerate}
 \item The variation of particle wind will result in a variation of $\dot{P}$, $\dot{P} \propto L_{\rm p}^{1/2}$. The may explain
the variation of $\dot{P}$ of many AXPs and SGRs (see Section 2.2 above).
A decreasing particle wind will result in a decreasing period derivative during magnetar outbursts
(Camilo et al. 2007; Levin et al. 2012; Anderson et al. 2012).


 \item Although magnetars and high magnetic field pulsars (HBPSRs) are close to each other on the
$P-\dot{P}$ diagram, they may be totally different from each other. In the case of wind braking,
magnetars are neutron stars with strong multipole field. While HBPSRs may be neutron stars only with strong dipole field.

Observationally, AXPs and SGRs have a larger level of timing noise (Gavriil \& Kaspi et al. 2002; Woods et al. 2002; Archibald et al. 2008).
This may be the result that
they are wind braking instead of magnetic dipole braking. Meanwhile, most HBPSRs do not show magnetar-like activities
which may be that most of them do not have as strong  multipole fields as magnetars (Ng \& Kaspi 2011; 
Pons \& Perna 2011).\footnote{If a HBPSR also has a strong multipole field,
it can also show magnetar-like activities. This may be the case of PSR J1846--0258 (Gravill et al. 2008; Pons \& Perna 2011).}
\end{enumerate}

\begin{figure}[!htbp]
 \centering
\includegraphics{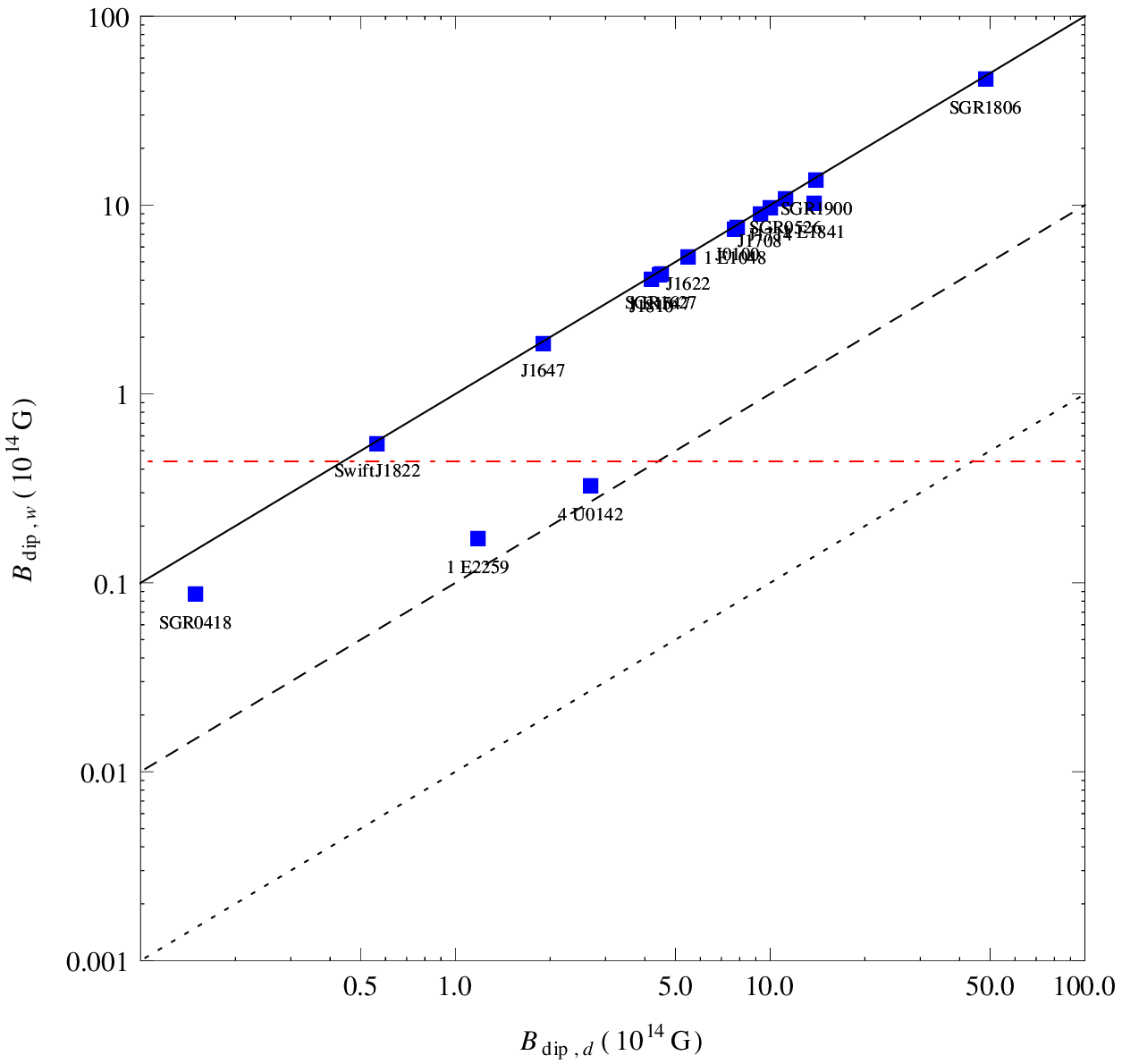}
\caption{The same as Figure \ref{fig_Bdip1}, assuming $\theta_{\rm s} =0.05$ and $L_{\rm p}=L_{\rm x}$. 
See text for details.}
\label{fig_Bdip4}
\end{figure}

The above calculations are done by assuming $L_{\rm w} =L_{\rm p}$. 
As discussed in Section \ref{section_Lw}, a strong reduction of dipole magnetic field is possible only when 
$L_{\rm w}$ is comparable to $L_{\rm p}$. This corresponds to a very collimated particle wind at the star surface. 
A small reduction of dipole magnetic field results from 
detailed modeling of wind luminosity. Assuming a constant polar cap opening angle 
$\theta_{\rm s} =0.05$ and $L_{\rm p} =L_{\rm x}$, 
the corresponding dipole magnetic field is also shown in table \ref{table_Bdip} and Figure \ref{fig_Bdip4}
(The case is similar when assuming $\theta_{\rm s} =0.05$ and $L_{\rm p} = 10^{35} \,\rm erg \,s^{-1}$). 
As pointed out in the beginning of this section, if the wind luminosity is lower than the rotational
energy loss rate, the dipole magnetic field will be the same as that of magnetic dipole braking. 
Then, the effect of a particle wind will mainly reflected in higher order modifications, e.g. 
braking index, timing noise, a magnetism-powered pulsar wind nubulae etc. 
Considering detailed modeling of wind luminosity,
the dipole magnetic field will be the same as the magnetic dipole braking case for most sources.
Only for four sources, their dipole magnetic fields are lower than the magnetic dipole braking case.

\begin{table}
\footnotesize
  \caption{Measured quantities and inferred dipole magnetic field of magnetars.}\label{table_Bdip}
  \begin{tabular}{llllllll}
    \tableline\tableline
    source name & $P$ & $\dot{P}$ & $L_{\rm x}$ & $B_{\rm dip,d}$
    & $B_{\rm dip,w}$ & $B_{\rm dip,w}$ & $B_{\rm dip,w}$\\
     & second & $10^{-11}$ & $10^{35}\, \rm erg\, s^{-1}$ & $10^{14} \,\rm G$ & $10^{14} \,\rm G$ & $10^{14} \,\rm G$ & $10^{14} \,\rm G$\\
    \tableline
    SGR 0526--66 & 8.05 & 3.8 & 1.4 & 11.2 & 1.9 & 1.6 & $B_{\rm dip,d}$\\
    SGR 1900+14 & 5.2 & 9.2 & 0.83-1.3$^a$ & 14.0 & 7.1 & 6.9 & $B_{\rm dip,d}$\\
    SGR 1806--20 & 7.6 & 75 & 1.6 & 48.3 & 39.5 & 31.2 & $B_{\rm dip,d}$\\
    SGR 1627--41 & 2.59 & 1.9 & 0.025 & 4.5 & 2.9 & $B_{\rm dip,d}$ & $B_{\rm dip,d}$\\
    SGR 0418+5729 & 9.08 & $<0.0006^{b}$ & $<0.00062^{b}$ & 0.15 & 0.00026 & 0.011 & 0.09\\
    Swift J1822.3-1606 & 8.44 & 0.0092 & 0.004 & 0.56 & 0.0044 & 0.069 & $B_{\rm dip,d}$\\
    \tableline
    4U 0142+61 & 8.69 & 0.203 & 1.1 & 2.7 & 0.093 & 0.089 & 0.34 \\
    1E 1048.1--5937 & 6.46 & 2.25 & 0.059 & 7.7 & 1.4 & 5.7 & $B_{\rm dip,d}$\\
    1E 2259+586 & 6.98 & 0.0484 & 0.34 & 1.2 & 0.028 & 0.048 & 0.18\\
    1E 1841--045 & 11.78 & 3.93 & 1.9 & 13.8 & 1.3 & 0.97 & 10.6\\
    1E 1547.0--5408 & 2.07 & 2.318 & 0.0058 & 4.4 & $B_{\rm dip,d}^{*}$ & $B_{\rm dip,d}$ & $B_{\rm dip,d}$\\
    1RXS J170849.0--400910 & 11.0 & 1.91 & 0.59 & 9.3 & 0.69 & 0.9 & $B_{\rm dip,d}$\\
    XTE J1810--197 & 5.54 & 0.777 & 0.00031 & 4.2 & 0.56 & $B_{\rm dip,d}$ & $B_{\rm dip,d}$\\
    CXOU J010043.1--721134 & 8.02 & 1.88 & 0.61 & 7.9 & 0.94 & 1.2 & $B_{\rm dip,d}$\\
    CXO J164710.2--455216 & 10.61 & 0.083 & 0.0044 & 1.9 & 0.031 & 0.47 & $B_{\rm dip,d}$\\
    CXOU J171405.7--381031 & 3.83 & 6.40 & 0.22 & 10.0 & 6.7 & $B_{\rm dip,d}$ & $B_{\rm dip,d}$\\
    PSR J1622--4950 & 4.33 & 1.7 & 0.0063 & 5.5 & 1.6 & $B_{\rm dip,d}$ & $B_{\rm dip,d}$\\
    \tableline
  \end{tabular}
\flushleft
\textbf{Notes:} Column one to eight are respectively, source name, period, period derivative, persistent X-ray luminosity
in the $2-10\,\rm keV$ range, dipole magnetic field assuming magnetic dipole braking, dipole magnetic field in the case of
wind braking assuming a wind luminosity $L_{\rm w} =L_{\rm p}=10^{35} \,\rm erg \,s^{-1}$, dipole magnetic field in the case of
wind braking assuming a wind luminosity $L_{\rm w} =L_{\rm p}=L_{\rm x}$, and dipole magnetic field in the case of wind braking 
assuming $\theta_{\rm s} =0.05$ and $L_{\rm p}=L_{\rm x}$. 
All data are from the McGill SGR/AXP online catalogue
(http://www.physics.mcgill.ca/$\sim$pulsar/magnetar/main.html, up to January, 27, 2012), except for SGR 0418+5729
(from Rea et al. 2010), and Swift J1822.3--1606 (from Rea et al. 2012a).
The first column are ordered roughly by the source's discover time. 

$^a$: median value is used during calculations

$^b$: upper limit is used during calculations

$^*$: When the wind luminosity is smaller than the rotational energy loss rate,
the dipole magnetic field in the case of wind braking is the same as that of magnetic dipole braking.
A magnetism-powered particle wind mainly results in a magnetism-powered pulsar wind nebula
and other high order modifications. 
See text for details.

\end{table}

\subsection{Acceleration potential}

Most of the electromagnetic emission of magnetars is thought to originate in the closed field line region 
(Thompson et al. 2002; Beloborodov \& Thompson 2007; Tong et al. 2010b).
Meanwhile, since the rotational energy is always present, we should also see some rotation-powered
activities in magnetars (Zhang 2003). The rotation-powered activities are almost inevitable
especially when we assume that AXPs and SGRs are also magnetic dipole braking as rotation-powered pulsars
(Tong et al. 2011). The acceleration potential in open field line regions characterizes this point
quantitatively.

The maximum acceleration potential in pulsar open field line regions is (Ruderman \& Sutherland 1975)
\begin{equation}
 \Phi_{\rm max} =\frac{B_0 r_0^2 \Omega}{2c} \left( \frac{R_{\rm pc}}{r_0} \right)^2.
\end{equation}
In the case of magnetic dipole braking $R_{\rm pc} = r_0 (r_0/R_{\rm lc})^{1/2}$,
the corresponding acceleration potential is
\begin{equation}
 \Phi_{\rm max} = \left( \frac32 \frac{-\dot{E}_{\rm rot}}{c} \right)^{1/2}.
\end{equation}
In the case of wind braking, the polar cap radius is given by eq.(\ref{Rpc_wind_braking}).
Although the polar cap radius is larger than the magnetic dipole braking case,
the dipole magnetic field is lower. The net effect will be concealed.
The corresponding acceleration potential is
\begin{equation}
 \Phi_{\rm max} = \left( \frac{\sqrt{3}}{2} \frac{-\dot{E}_{\rm rot}}{c} \right)^{1/2}.
\end{equation}
The maximum acceleration potential is the same (within a factor of two) in the wind braking case
and magnetic dipole braking case.

Although the maximum acceleration potential is the same, the detailed acceleration mechanism will
be qualitatively different. In the presence of a particle wind, vacuum gaps may not be formed, e.g.,
outer gap etc. This may explain the conflicts between outer gap model in the case of magnetars and {\it Fermi}
observations (Tong et al. 2010a, 2011). Meanwhile, space charge limited flow type acceleration mechanism
may still exist (Xu 2007). In a wind loaded magnetosphere, detailed calculations of space
charge limited flows are needed in the future.

In calculating Figure \ref{fig_Bdip2}, we show that only those sources with $L_{\rm x}>-\dot{E}_{\rm rot}$
are wind braked down. While for sources with $L_{\rm x}<-\dot{E}_{\rm rot}$, they are still magnetic dipole braking, the
same as rotation-powered pulsars. A magnetosphere similar to that of rotation-powered pulsars
is prepared during the persistent state. This may be taken as the initial state. An outburst will may trigger
the radio emission of magnetars as observed. Then it is natural that only sources with $L_{\rm x}<-\dot{E}_{\rm rot}$ can have
radio emissions. This may explain the ``fundamental plane'' of magnetar radio emissions found by Rea et al. (2012b).
More detailed investigations are needed.

\subsection{Spin down evolution and age}

A given magnetar, with dipole magnetic field $B_0 =b_0\times B_{\rm QED} =b_0\times 4.4\times 10^{13} \,{\rm G}$
($b_0 \sim 1$ from eq.(\ref{Bdipw2})),
and a wind luminosity $L_{\rm w} =L_{\rm p} =L_{\rm p,35} \times 10^{35} \,{\rm erg\,s^{-1}}$, 
will evolve from magnetic dipole braking at early stage to wind braking at later stage. 
At present, we assume that $B_0$ and $L_{\rm p}$ are both
constants (A decaying particle wind will be considered in Section 5.1 below). From eq.(\ref{LpEdotD}) and eq.(\ref{Edot_D}),
at the early stage, the star rotates very fast and $\dot{E}_{\rm{d}}$ is larger than $L_{\rm p}$. Therefore,
the star will be braked down by magnetic dipole radiation at early stage. However, at later stage,
the star will be slowed down and $\dot{E}_{\rm{d}}$ will be smaller than $L_{\rm p}$. Therefore, the star
will become wind braking at later stage. The transition from magnetic dipole braking to wind braking happens
at
\begin{equation}
L_{\rm p} =\dot{E}_{\rm{d}} =\frac{B_0^2 r_0^6 \Omega^4}{6 c^3}.
\end{equation}
The corresponding rotation period is
\begin{equation}\label{P1}
 P_1 =0.66 \,b_0^{1/2} L_{\rm p,35}^{-1/4} \,{\rm s}.
\end{equation}
$P_1$ can also be obtained by requiring $r_{\rm open} \leq R_{\rm lc}$ (Thompson et al. 2000).
When the star's rotation period is less than $P_1$ it will be braked down by magnetic dipole radiation.
The corresponding period derivative at the transition point is
\begin{equation}
 \dot{P}_1 =7.2\times 10^{-13} \,b_0^{3/2} L_{\rm p,35}^{1/4}.
\end{equation}
If the magnetar rotation period at birth is much less than $P_1$, then the star age at $P_1$ is
\begin{equation}
 t_1 =\tau_{\rm c,1} \equiv \frac{P_1}{2 \dot{P}_1} =1.4\times 10^4 \,b_0^{-1} L_{\rm p,35}^{-1/2} \,{\rm yr}.
\end{equation}

The transition age $t_1$ is similar to the supernova remnant age associated with AXPs (Vink \& Kuiper 2006).
Beginning from $t_1$, $L_{\rm p} > -\dot{E}_{\rm rot}$ (eq.(\ref{validity})), the star will be braked down by a particle wind.
Furthermore, the particle wind of magnetars is from magnetic energy decay
$L_{\rm p} \sim -\dot{E}_{\rm B}$, where $E_{\rm B}$ is the star's magnetic energy
stored mainly in the form of multipole field. Therefore, during wind braking phase, $-\dot{E}_{\rm B} > -\dot{E}_{\rm rot}$.
The star's activities will be dominated by magnetic energy output rather than rotational energy output.
AXP/SGR-like activities may appear, i.e., the pulsar becomes a magnetar.

We now consider how a magnetar evolves from $(P_1,\dot{P}_1)$ to $(P_2,\dot{P}_2)$ (Thompson et al. 2000).
When we assume $B_0$ and $L_{\rm p}$ are both constants, then from eq.(\ref{Bdipw1}),
at wind braking phase
\begin{equation}
 \frac{\dot{P}}{P} =\frac{\dot{P}_1}{P_1} =\frac{\dot{P}_2}{P_2} = {\rm constant}.
\end{equation}
The period will evolve with time as
\begin{equation}\label{P_evolution}
 P_2 =P_1 \exp \{ \frac{t_2-t_1}{2 \tau_{\rm c,1}} \},
\end{equation}
where $t_2$ and $t_1$ are the star's true age at $P_2$ and $P_1$, respectively. $\tau_{\rm c,1}$ is the
characteristic age at $P_1$. For transition from magnetic dipole braking to wind braking, $t_1 =\tau_{\rm c,1}$.
However, in the general case, $t_1$ is not always equal to $\tau_{\rm c,1}$.
The star's age at a given period $P_2$ is
\begin{equation}
 t_2 = t_1 +2 \tau_{\rm c,1} \log \frac{P_2}{P_1}.
\end{equation}
After $t_1$, the star's period increases exponentially. For $P_2$ not very
large than $P_1$, we have $t_2\sim t_1 = \tau_{\rm c,1} =\tau_{\rm c,2}$, where $\tau_{\rm c,2}$
is the star's characteristic age at $P_2$.

\subsection{Braking index}

The braking index of a pulsar is defined as (Shapiro \& Teukolsky 1983)
\begin{equation}
 \dot{\Omega} = - ({\rm constant}) \Omega^n,
\end{equation}
where $n$ is called the braking index. $n=3$ for magnetic dipole braking.
For wind braking, from eq.(\ref{Bdipw1}), we have
\begin{equation}
 -I \Omega \dot{\Omega} = \left( \frac{B_0^2 r_0^6 \Omega^4}{6 c^3} \right)^{1/2} L_{\rm p}^{1/2}.
\end{equation}
Therefore, $n=1$ for wind braking (assuming $B_0$ and $L_{\rm p}$ are both constants).
The braking index of PSR J1734-3333, $n=0.9 \pm 0.2$, may imply that it is of wind braking
(Espinoza et al. 2011; a rotation-powered particle wind).
Future braking index measurement of a magnetar will help us make clear whether magnetars are magnetic dipole
braking or wind braking. Because the braking index will deviate from one if $B_0$ and/or $L_{\rm p}$ changes
with time, a braking index of a magnetar may also tell us the evolution
of its particle wind.

\subsection{Duty cycles of particle wind}

Harding et al. (1999) considered the duty cycles of a particle wind whose luminosity is
$L_{\rm p} \sim 10^{37} \,{\rm erg \,s^{-1}}$. It is shown that, due to the duty cycles
of particle wind, the dipole magnetic field and age vary continuously from
the dipole braking case to the wind braking case (Figure 1 in Harding et al. 1999).
However, the particle luminosity considered by Harding et al. (1999) is much stronger than
we considered here $L_{\rm p} \sim 10^{35} \,{\rm erg \,s^{-1}}$. It is possible that there
are two types of particle wind:
\begin{enumerate}
 \item A persistent component associated with the magnetar's persistent
emissions. The particle luminosity is $L_{\rm pp} \sim L_{\rm x} \sim 10^{35} \,{\rm erg\,s^{-1}}$.
 \item A burst component associated with outbursts of magnetars. The corresponding
particle luminosity may be about $L_{\rm pb} \sim L_{\rm burst} \sim 10^{37} \,{\rm erg \,s^{-1}}$.
\end{enumerate}
The burst component of particle wind may contribute to the enhanced spindown of magnetars after glitches
(Kaspi et al. 2003) and the possible ``radiation braking'' during giant flares of SGR 1900+14
(Thompson et al. 2000; Parfrey et al. 2012).

The long term averaged spindown of magnetars can be modeled similarly to that of Harding et al. (1999)
\begin{equation}
 -\langle \dot{E}_{\rm rot} \rangle = \dot{E}_{\rm w,burst} D_{\rm p} +\dot{E}_{\rm w,persistent} (1-D_{\rm p}),
\end{equation}
where $D_{\rm p}$ is the duty cycle of the burst component of particle wind.
From eq.(\ref{Edot_w}), the above equation can be rewritten as
\begin{equation}
 -\langle \dot{E}_{\rm rot} \rangle
=\dot{E}_{\rm{d}}^{1/2} L_{\rm pb}^{1/2} D_{\rm p} +\dot{E}_{\rm{d}}^{1/2} L_{\rm pp}^{1/2} (1-D_{\rm p})
=\dot{E}_{\rm{d}}^{1/2} L_{\rm eff}^{1/2},
\end{equation}
where $L_{\rm eff}^{1/2} =L_{\rm pb}^{1/2} D_{\rm p} +L_{\rm pp}^{1/2} (1-D_{\rm p})$ is the effective
particle luminosity. For typical parameters, the effective particle luminosity is
\begin{equation}
 L_{\rm eff,35}^{1/2} =10 \,L_{\rm pb,37}^{1/2} D_{\rm p} +L_{\rm pp,35}^{1/2} (1-D_{\rm p}).
\end{equation}
For $D_{\rm p}=0$, this is just the case we considered above. For $D_{\rm p}=1$, this corresponds
to a strong wind case ($L_{\rm p} =10^{37}\,{\rm erg\, s^{-1}}$) as considered by Harding et al. (1999).
The duty cycle can be estimated from the observations of the transient magnetar SGR 1627-41 (Mereghetti et al. 2009).
The duration between two outbursts is about ten years. Therefore, the maximum value of duty cycle is about $0.1$.
The corresponding effective particle luminosity is
$L_{\rm eff,35}^{1/2} =L_{\rm pb,37}^{1/2} + 0.9 L_{\rm pp,35}^{1/2}$, about two times larger than the persistent
component of the particle wind. In conclusion, the previous discussions are still valid considering the
possible existence of a burst component of particle wind.

\section{Discussions}

\subsection{A decaying particle wind}

In the magnetar model, both the persistent and burst emissions of AXPs and SGRs
are powered by magnetic field decay. The total magnetic field will
decay with time. Meanwhile, the photon luminosity as well as the particle luminosity
will also evolve with time. Eventually both the photon luminosity and particle
luminosity will also decay with time (Turolla et al. 2011). In the case of a decaying particle wind,
the spin down evolution of magnetars will be different from previous considerations.
Considering different avenues for magnetic field decay, the total magnetic field may decay
with time in a power law form (Heyl \& Kulkarni 1998). The consequent magnetic energy decay
rate $-\dot{E}_{\rm B}$ will also of power law form. Since the particle luminosity is
from the magnetic energy decay, we may assume a power law form of particle luminosity
\begin{equation}
 L_{\rm p}(t) =L_{\rm p,0} \left( \frac{t}{t_{\rm D}} \right)^{-\alpha}, \ 0\le \alpha \le 2,
\end{equation}
where $L_{\rm p,0}$ and $\alpha$ are constants, $t_{\rm D}$ is the time when the magnetic field
starts to decay significantly. $t_{\rm D}$ may be of the same order as $t_1$ when wind braking starts to operate.
For $\alpha$ larger than two, $L_{\rm p}(t)$
decays more rapidly than $\dot{E}_{\rm{d}}$. In this case, the wind braking criterion is not fulfilled
(eq. (\ref{LpEdotD})).
In the case of decaying particle wind,
by integrating eq. (\ref{Bdipw1}), we can get the spin down evolution of magnetars.
For $0 \le \alpha < 2$, the period evolves with time
\begin{equation}
 P_2 =P_1 \, \exp\{ \frac{t_2 (t_2/t_1)^{-\alpha/2}-t_1}{(2-\alpha) \tau_{\rm c,1}} \},
\end{equation}
and the star age at a given period is
\begin{equation}
 t_2 \left( \frac{t_2}{t_1} \right)^{-\alpha/2} =t_1 + (2-\alpha) \tau_{\rm c,1} \log \frac{P_2}{P_1}.
\end{equation}
For the special case of $\alpha=2$, the corresponding expressions for period and age are
\begin{equation}\label{Palpha2}
 P_2 =P_1 \, \left( \frac{t_2}{t_1} \right)^{t_1/2\tau_{\rm c,1}} ,
\end{equation}
\begin{equation}
t_2  = t_1 \left( \frac{P_2}{P_1} \right)^{2\tau_{\rm c,1}/t_1}.
\end{equation}
Equation (\ref{Palpha2}) is now the same as the magnetic dipole braking case by setting
$t_1 =\tau_{\rm c,1}$ (eq. (5.18) in Lyne \& Graham-Smith 2012).

\subsubsection{Calculation of braking index}

\begin{figure}[!htbp]
 \centering
\includegraphics{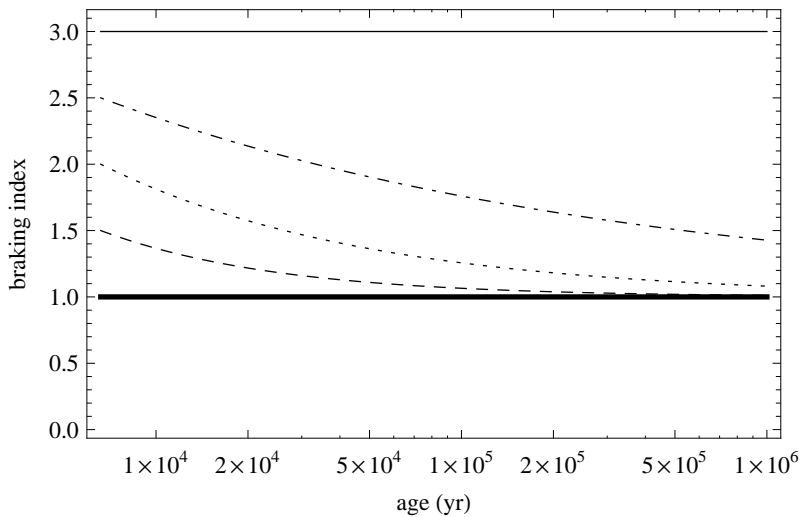}
\caption{Braking index in the case of wind braking as a function of age.
The parameters of AXP 4U 0142+61 are used.
The thick solid, dashed, dotted, dot-dashed, and thin solid lines are for
$\alpha =0,\ 0.5,\ 1,\ 1.5,\ 2$, respectively.}
\label{n0142}
\end{figure}

The braking index predicted for the most luminous AXP (4U 0412+61, Dib et al. 2007) is 
shown as function of age in Figure \ref{n0142}. $L_{\rm p,0} =10^{37} \, \rm erg\, s^{-1}$ is assumed.
For a constant particle wind, the braking index $n=1$ is obtained, as previously discussed.
For the critical case $\alpha=2$, the braking index $n=3$ is obtained the same as the magnetic dipole
braking case, as can be seen from eq. (\ref{Palpha2}). For the intermediate case
$0<\alpha<2$, a braking index $n=1-3$ is obtained. Future braking index measurement of this source
may tell us whether it is wind braking or magnetic dipole braking.

\subsection{The presence of a fallback disk}

In the case of wind braking, the star's true age is of the same order of the characterristic age
$t \sim \tau_{\rm c}$.
For those magnetars whose supernova remnant age $t_{\rm snr} \sim \tau_{\rm c}$, then it is understandable
that they are wind braking. However, for AXP 1E 2259+586, its supernova remnant age
$t_{\rm snr} \approx 10^{4} \,{\rm yr} \ll \tau_{\rm c}=23\times 10^{4} \,{\rm yr}$ (Vink \& Kuiper 2006).
For a decaying particle wind, the star's true age can be less than $\tau_{\rm c}$. However,
$L_{\rm p}(t_{\rm snr})$ will be larger than $L_{\rm x} \sim 10^{35} \,\rm erg \,s^{-1}$. Therefore,
additional torque may be needed for AXP 1E 2259+586.

\begin{figure}[!htbp]
 \centering
\includegraphics{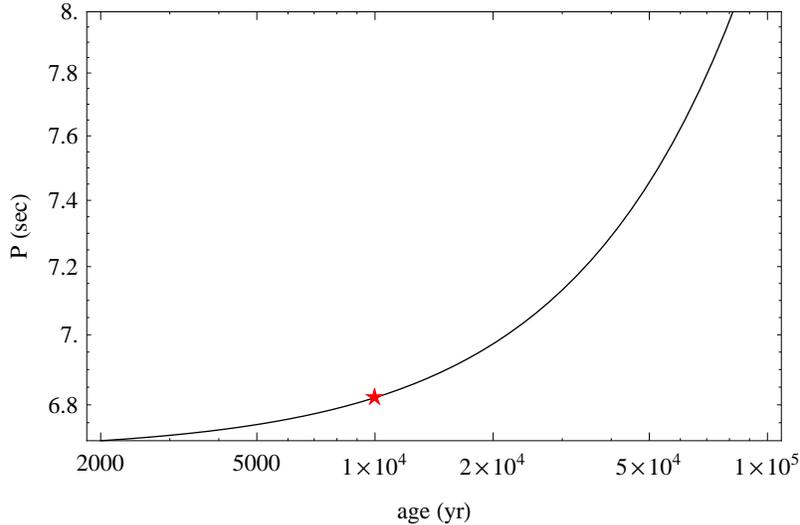}
\caption{Evolution of rotation period as a function of age, calculations for AXP 1E 2259+586.
The star is AXP 1E 2259+586, $t_{\rm snr}$ is taken as the true age.}
\label{age2259}
\end{figure}

The presence of a fallback disk may help to solve this age discrepancy
(Shi \& Xu 2003). At early phase, a fallback disk provides the braking torque of the magnetar.
At the end of disk braking, the star has been slowed down significantly, e.g. $t_1 = 2\times 10^{3} \,{\rm yr}$, $P_1=6.7 \,\rm s $.
For a particle luminosity $L_{\rm p} =10^{35} \,\rm erg\, s^{-1}$,
the evolution of rotation period is shown in Figure \ref{age2259}.

Observationally, there may be a debris disk around 1E 2259+586 (Kaplan et al. 2009).
If we assume that SGR 0418+5729 is also a young magnetar, then a fallback disk is also needed (in the early stage)
to spin down it to the present period (Alpar et al. 2011). For the disk torque to operate effectively,
the dipole magnetic field can not be too high, e.g., $B_{\rm dip}=10^{12}-10^{13} \,\rm{G}$
is required (Shi \& Xu 2003; Alpar et al. 2011). This is consistent with the dipole magnetic field
obtained by assuming wind braking (see eq.(\ref{Bdipw2})).

In conclusion, there may be a fallback disk in the early stage of a magnetar.
This fallback disk may help to solve the age discrepancy.
At present, they have been slowed down significantly and have become wind braking.

\subsection{Spin down evolution of newly born magnetars}

Magnetars are thought to be descendants of rapidly rotating proto-neutron stars, with rotation
period $\sim 1 \,\rm ms$ (Duncan \& Thompson 1992). A strong dipole field ($B_{\rm dip} \sim 10^{15} \,\rm G$)
will cause the spin down time scale of the magnetar less than the supernova shock breakout time.
This will cause the supernova associated with magnetar birth more energetic (Duncan \& Thompson 1992).
However, studies of supernova remnants associated with AXPs and SGRs show that the putative
supernova energies are of canonical value (Vink \& Kuiper 2006). This provides challenges
to the traditional magnetar model. If magnetars are wind braking instead of
magnetic dipole braking, then they will have much weaker dipole field. The corresponding spin down
time scale will be much longer than the shock breakout time,
$\tau_{\rm sd} \sim 60 B_{14}^{-2} (P_{\rm i}/1 \, {\rm ms})^2 \, \rm hr$. This may explain the observations of
Vink \& Kuiper (2006).

Moreover, in the presence of strong multipole field, magnetars are of prolate shape. This may cause them
to emit strong gravitational waves after birth (Dall'Osso et al. 2009). The gravitational waves will also carry
away some amount of the initial rotational energy. For gravitational wave to operate effectively, its competing
process (i.e., magnetic dipole braking) can not be too strong. Therefore, a weaker magnetic dipole field is required
$B_{\rm dip} \lesssim 10^{14} \,\rm G$. This is also consistent with the result of wind braking. In the actual case,
a combination of these two processes, i.e., longer spin down time scale and gravitational wave emissions, may account
for the observations. Their contributions depends on the dipole and multipole field strength
of the star, which may vary from source to source.

\subsection{Magnetism-powered pulsar wind nebula}

A particle wind with luminosity $10^{35} \, \rm erg\, s^{-1}$ may produce a visible
nebula around the central magnetar (The putative nebula may also contain contributions from a rotation-powered
particle wind). This pulsar wind nebula is magnetism-powered in nature
since the particle wind is originated from magnetic field decay. There may be a pulsar wind nebula
around AXP 1E 1547.0-5408 (Vink \& Bamba 2009). Since both the particle wind and the persistent
X-ray luminosity of a magnetar are from magnetic field decay, there will be a strong correlation
between them. Then for a magnetism-powered pulsar wind nebula, we should see a correlation between
the nebula luminosity and the magnetar luminosity. This is just the case of Figure 2 in Olausen et al. (2011).
In Olausen et al. (2011), they see a strong correlation between the extended emission of AXP 1E 1547.0-5408
and its source flux. Therefore, Olausen et al. concluded that the pulsar wind nebula origin for the
extended emission is ruled out and it is a dust scattering halo. However, a strong correlation between the
extended emission and the source flux just rules out the rotation-powered pulsar wind nebula hypothesis.
Such a correlation is a natural result if the pulsar wind nebula is magnetism-powered.
Future multiband observations of this source may tell us whether it is a magnetism-powered
pulsar wind nebula or a dust scattering halo.

For a magnetism-powered pulsar wind nebula, an extreme case is that the nebula luminosity can exceed that of the
star's rotational energy loss rate $L_{\rm pwn}>-\dot{E}_{\rm rot}$. However,
for young magnetars, their rotational energy loss rates are also very
high. Therefore, the extreme case may be very hard to achieve.
A possible case is that we can see a high conversion efficiency of the putative nebula. The possible pulsar wind nebula seen
around RRAT J1819$-$1458 has a relatively high conversion efficiency (Rea et al. 2009).
It may contain contributions from a magnetism-powered particle wind.

\section{Conclusions}

Considering recent observations challenging the traditional magnetar model (neutron stars with both
strong dipole field and strong multipole field), we explore the
wind braking of magnetars. There are some observational clues for the existence of a magnetism-powered
particle wind. The total particle luminosity is estimated to be $\sim 10^{35} \,\rm erg \, s^{-1}$,
comparable to their persistent X-ray luminosities. Such a particle wind will amplify the star's rotational energy
loss rate. The consequent dipole magnetic field is about ten times smaller than that of magnetic dipole braking,
if the particle flow is strongly collimated at the star surface. In the wind braking scenario, 
magnetars are neutron stars with strong multipole field. For some sources, a strong dipole field may be no longer necessary.  
Wind braking of magnetars may help us to explain some challenging observations of magnetars.

A magnetism-powered pulsar wind nebula and a braking index smaller than three are the two
predictions of the wind braking model\footnote{After we put this paper on the arXiv(1205.1626),
Tendulkar et al. (arXiv:1210.8151) provide some indirect information of braking index of magnetars.
Their information is consistent with our analysis here (a braking index smaller than three).}.
Future studies will tell us whether magnetars are wind braking or
magnetic dipole braking.

\acknowledgments

The authors would like to thank the Referee for detailed and thoughtful comments,
and B. Zhang for helpful discussions. 
H. Tong would like to thank KIAA at PKU for support of visiting.
This work is supported by National Basic Research Program of China (2012CB821800, 2009CB824800),
National Natural Science Foundation of China (11103021, 11225314, 10935001, 10833003),
West Light Foundation of CAS (LHXZ201201), and the John Templeton Foundation.



\end{document}